\documentclass[aip,reprint]{revtex4-1}
\usepackage{graphicx}
\usepackage{dcolumn}
\usepackage{bm}
\usepackage[mathlines]{lineno}

\usepackage[utf8]{inputenc}
\usepackage[T1]{fontenc}
\usepackage{mathptmx}
\usepackage{etoolbox}
\usepackage{amsmath}
\usepackage{booktabs}
\usepackage[colorlinks=true,
            linkcolor=cyan,
            citecolor=cyan,
            urlcolor=cyan]{hyperref}

\usepackage[table]{xcolor}
\usepackage{lineno}
\usepackage{graphicx}
\usepackage{epsfig}
\usepackage{svg}
\draft 

\begin{document}

\makeatletter
\def\@email#1#2{%
 \endgroup
 \patchcmd{\titleblock@produce}
  {\frontmatter@RRAPformat}
  {\frontmatter@RRAPformat{\produce@RRAP{*#1\href{mailto:#2}{#2}}}\frontmatter@RRAPformat}
  {}{}
}%
\makeatother

\title{Scaling of nonlinear dynamics driven by stimulated Raman scattering in gas-filled hollow-core fibers} 

\author{P. Arcos}
 \affiliation{Department of Communications Engineering, University of the Basque Country (UPV/EHU), Torres Quevedo 1, 48013 Bilbao, Spain}
 \email{pau.arcos@ehu.eus}

\author{A. Mena}
 \affiliation{Department of Communications Engineering, University of the Basque Country (UPV/EHU), Torres Quevedo 1, 48013 Bilbao, Spain}
 
 \author{M. S\'anchez-Hern\'andez}
 \affiliation{Department of Communications Engineering, University of the Basque Country (UPV/EHU), Torres Quevedo 1, 48013 Bilbao, Spain}
 
\author{A. Berganza}%
 \affiliation{Department of Applied Mathematics, University of the Basque Country (UPV/EHU), Torres Quevedo 1, 48013 Bilbao, Spain}

\author{B. Garcia-Ramiro}%
 \affiliation{Department of Applied Mathematics, University of the Basque Country (UPV/EHU), Torres Quevedo 1, 48013 Bilbao, Spain}

\author{J. Zubia}%
 \affiliation{Department of Communications Engineering, University of the Basque Country (UPV/EHU), Torres Quevedo 1, 48013 Bilbao, Spain}%
 \affiliation{EHU Quantum Center, University of the Basque Country (UPV/EHU), 48013 Bilbao, Spain}%
  
\author{D. Novoa}%
 \affiliation{Department of Communications Engineering, University of the Basque Country (UPV/EHU), Torres Quevedo 1, 48013 Bilbao, Spain}
\affiliation{EHU Quantum Center, University of the Basque Country (UPV/EHU), 48013 Bilbao, Spain}%
\affiliation{IKERBASQUE, Basque Foundation for Science, Plaza Euskadi 5, 48009 Bilbao, Spain}%

\date{\today}

\begin{abstract}
Optical systems are scalable under low-intensity illumination since their governing equations are linearly dependent of the optical signal strength. Nonetheless, in high-intensity regimes, the induced polarization becomes nonlinear, rendering the simple scalability of the previous systems invalid. Despite this, canonical nonlinear phenomena such as filamentation and high-harmonic generation in free space have recently been demonstrated to be scalable. Here we will discuss the extension of the scale-invariance paradigm to stimulated Raman scattering and molecular modulation in hollow anti-resonant fibers filled with Raman-active gases. We have found that the complex in-fiber dynamics can be accurately reproduced under very different conditions by keeping the so-called gain reduction factor, that accounts for the coupling of the interacting fields, as well as the dephasing time $T_2$ unaltered. Such scaling strategy enables access to equivalent nonlinear propagation scenarios without sacrificing performance, laying the foundations for to the design of nonlinear devices operating in exotic frequencies, like the ultraviolet, or quantum frequency convertors of non-classical light.
\end{abstract}


\maketitle 

\section{Introduction}
Linear optical systems are inherently scalable since the solutions of Maxwell’s equations in the low intensity regime are linearly dependent of the optical signal strength. In sharp contrast, when high light intensities are considered and the polarization induced in the optical medium becomes nonlinear, such simple scaling is no longer valid and different input intensities will, in general, yield totally different propagation dynamics~\cite{Boyd}. It has recently been recognized, however, that judicious adjustment of various parameters of the system such as propagation distance, light intensity and beam size, enables the scalability of highly nonlinear phenomena such as filamentation or high-harmonic generation~\cite{Heyl:16,rothhardt2014absorption}. On the other hand, specialty waveguides such as gas-filled hollow-core fibers guiding broadband through anti-resonant reflection (ARFs) are excellent platforms to observe such scaling. This is due to the exquisite control they offer over the various physical parameters affecting nonlinear propagation of intense optical pulses, as recently demonstrated in the context of soliton self-compression in noble gases~\cite{Schade:21}.

The situation becomes increasingly more complex when Raman-active molecular gases are considered. In stimulated Raman scattering (SRS), a pump pulse launched in the ARF is inelastically scattered off the gas molecules confined in the core, getting down-shifted by the corresponding Raman transition frequency $\Omega_R$ to a Stokes band and triggering the excitation of a coherence wave (or optical phonon) of synchronized molecular motion~\cite{arcos2024narrowband}. These coherence waves then favor an exponential amplification of the Stokes signal and, provided specific phase-matching conditions are fulfilled, can also promote the frequency up-conversion by $\Omega_R$ of further pump photons (or any other arbitrary signal) to the anti-Stokes band through molecular modulation~\cite{Bauerschmidt:15}. In contrast to quasi-instantaneous nonlinear phenomena, it turns out that SRS-driven interactions are not straightforwardly scalable using the rules derived in Refs.~\onlinecite{Heyl:16,Schade:21}. This is because they do not capture the influence on the dynamics of the inherent coupling of the different bands through the induced polarization~\cite{PhysRevLett.12.504}, as well as the temporal nonlocality of these effects within the lifetime of the induced coherence~\cite{agrawal2000nonlinear}. Therefore, since SRS lies at the heart of many fiber-based applications, a specific scaling strategy capable of providing different sets of physical parameters yielding equivalent nonlinear SRS dynamics in gas-filled ARFs would be desirable.

In this article we will discuss the nontrivial extension of the scale-invariance paradigm to SRS and molecular modulation in gas-filled ARFs. By keeping constant the gain reduction factor $\rho$~\cite{bauerschmidt2015dramatic} that characterizes the nonlinear interplay among the interacting fields, as well as the dephasing time $T_2$ of the molecular oscillations, we can reproduce the in-fiber dynamics with high fidelity within realistic constraints. This is achieved by using mixtures of Raman-active and noble gases, and tailoring both core radius and capillary-wall thickness to obtain equivalent nonlinear evolution with different parameters. Our results pave the way to the design and optimization of fiber-based nonlinear devices such as broadband\cite{gao2022raman,belli2015vacuum} and narrowband light sources operating in exotic frequency domains like the ultraviolet or the mid-infrared ~\cite{tyumenev2020narrowband,belli2015vacuum,mridha2019thresholdless,wang2020high,li20203,wang2024synthesizing,Gladyshev:24}, or quantum frequency convertors of non-classical light~\cite{tyumenev2022tunable,hamer2024frequency}.

\section{Theoretical model}\label{theory}

The nonlinear propagation dynamics driven by SRS in gas-filled ARFs pumped by narrowband pulses can be accurately modeled using a single-mode set of coupled Maxwell-Bloch equations\cite{Bauerschmidt:15}:
\begin{eqnarray}
    \frac{\partial E_l}{\partial z} &=& - i\kappa_{2,l}\frac{\omega_l}{\omega_{l-1}}QE_{l-1}s_{l-1}s_l^* \nonumber\\ &&-i\kappa_{2,l+1}Q^*E_{l+1}s_{l+1}s_{l}^*-\frac{1}{2}\alpha_lE_l,\label{eq:MBe_equation00}\\
     \frac{\partial Q}{\partial t} + \frac{Q}{T_2} 
     &=& - \frac{i}{4}\sum_l\kappa_{1,l}E_lE_{l-1}^*s_ls_{l-1}^*, 
     \label{eq:MBe_equation0}
\end{eqnarray}
where $Q$ is the amplitude of the molecular coherence and $E_l$ is the electric-field envelope of the $l$ sideband of central frequency $\omega_l = \omega_p + l\cdot \Omega_R$, with $\omega_p$ being the pump frequency. Here, $l=-1, 0, 1$ would represent the Stokes (S), pump (P) and anti-Stokes (AS) frequencies respectively. Without loss of generality, our analysis will be focused on the fundamental vibrational transition in hydrogen ($\Omega_R$ $\approx$ 125 THz), although it could be straightforwardly extended to any other Raman-active gases or transitions. The spatial phase progression is described by the term $s_{l} = \exp(-i\beta_{l}z)$, where $\beta_l$ is the propagation constant of the sideband $l$. The coupling constants are $\kappa_{1,l}=(2g_lc^2\epsilon_0^2/N\hbar(\omega_l-\Omega_R)T_2)^{1/2}$ and $\kappa_{2,l}= N\hbar(\omega_l-\Omega_R)\kappa_{1,l}/2\epsilon_0c$, $N$ is the molecular number density, $c$ is the speed of light in vacuum, $\hbar$ is the reduced Planck's constant, $\epsilon_0$ is the vacuum permittivity and 
\begin{equation}
    g_l=\frac{9.37\times10^{12}(57.2p_R/\Delta\nu)(\omega_l-\Omega_R)}{c(7.19\times10^{13}-\omega_l^2/c^2)^2}
    \label{eq:ramangain}
\end{equation}
is the vibrational Raman gain in hydrogen\cite{bischel1986wavelength}, being $p_R$ the partial pressure of the Raman-active gas. The second term in Eq.~\eqref{eq:MBe_equation0} is phenomenologically introduced to account for the dephasing of the molecular coherence and is characterized by the time $T_2=1/\pi\Delta\nu$, where the Raman linewidth $\Delta\nu$ can be approximated in the general case of a binary mixture of Raman-active and inactive buffer gases as:
\begin{equation}
    \Delta\nu = \frac{A(p_R,p_B)}{p_R}+Bp_R+Cp_B.
    \label{eq:Raman_linewidth}
 \end{equation}
 In the latter, $p_B$ is the partial pressure of the Raman-inactive buffer gas (xenon in this work, as motivated by recent experiments\cite{hosseini2017enhanced}), which can be used to increase the Raman linewidth through collisions with the molecules and is an important ingredient of our scaling strategy, as it will be discussed below. The parameter $A$ depends on the self-diffusion coefficients of the different gases and the Raman shift\cite{difusion_multigas, Raman_linweidths}, $B=48$ MHz/bar and $C=380$ MHz/bar for a H$_2$-Xe mixture\cite{hosseini2017enhanced}. 
 
\begin{figure}[h!]
\includegraphics{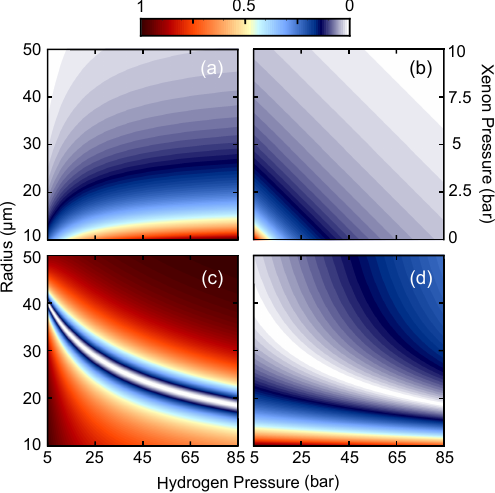}
    \caption{Representation of the (a) inverse gain length $g_0I_p$, (b) dephasing time $T_2$, (c) gain reduction factor $\rho$ and (d) the absolute value of the dephasing $\vartheta$ as function of the fiber core radius and filling gas pressure. The filling gas in (a),(c) and (d) is pure H$_2$, whereas in (b) we have also considered the influence of xenon as buffer gas. Total gain suppression $\rho=0$ occurs at $\vartheta=0$ (white-shaded areas in (c),(d)). All quantities are normalized to their own maxima. For these calculations, we assume a capillary-wall thickness of 270 nm, a pump energy of 100 \textmu J and 3.8 ns-long, 1064 nm-pump pulses.}
    \label{FIG1}
\end{figure}

 In Fig.~\ref{FIG1}(a) we represent the product $g_0I_p$ where $I_p$ is the intensity of the pump pulse {(unless otherwise stated, we will consider 3.8 ns-long, 1064 nm pump pulses, inspired by the system used in Ref. \onlinecite{tyumenev2022tunable})} and $\cal E$ = 100 \textmu J is its energy. As it can be seen, the vibrational Raman gain in hydrogen tends to saturate above $\sim$10 bar due to the simultaneous increase of both $p_R$ and $\Delta\nu$ with $N$. In Fig.~\ref{FIG1}(b) we show the dependence of the decoherence time $T_2$ with the partial pressures of both Raman-active and buffer gases, highlighting the effect of the collisional dephasing as the pressure of xenon increases.

On the other hand, the propagation constant of the light coupled to the fundamental core mode of an ARF can be approximated with good accuracy (in absence of loss-inducing resonances with modes localized in the glass walls of the tubular cladding) by the so-called Marcatili-Schmeltzer model \cite{Marcatilli}:
\begin{equation}
    \beta_l= \sqrt{(\omega_l n_g/c)^2-(u_{01}/a)^2},
    \label{eq:propagationctt0}
\end{equation}

where $n_g(\omega_l,p) = \sqrt{1+\delta(\omega_l)\tilde p\tilde T}$ is the refractive index of the filling gas, parametrized through a Sellmeier relation $\delta(\omega_l)$ \cite{peck1977refractivity, borzsonyi2008dispersion}, and the normalized pressure $\tilde p =p/p_0$ ($p_0=1$ bar), and temperature $\tilde T =T_0/T$ ($T_0=273.15$ K). We will consider room temperature $T=298$ K in all cases under study. The second term in Eq.~\eqref{eq:propagationctt0} includes the effective modal radius $a \approx 1.07r$ \cite{wei2017negative}, where $r$ is the fiber core radius, and $u_{01}=2.4048$ (i.e. the first root of the zero order Bessel function of the first kind).  Finally, although we will use the resonance-free dispersion model represented by Eq.~\eqref{eq:propagationctt0} to derive our analytical scaling rules, the full simulations will be performed using a more refined resonance-capturing model \cite{zeisberger2017analytic} that will be described in more detail at the end of the work.

\section{Scaling of nonlinear SRS dynamics}\label{scaling principles}

The scaling strategy for nonlinear optical phenomena introduced in the seminal work by Heyl \textit{et al.} \cite{Heyl:16} was originally applied to diluted gas-based free-space geometries irradiated by ultrashort pulses. Our aim is to follow the same spirit to scale the nonlinear dynamics occurring during the interaction of guided narrowband pulses with micro-confined molecular gases over a much wider range of pressures, including the role of gas mixtures and the inherent optical properties of ARFs. This is very relevant, e.g., the design of new fiber-based Raman lasers, frequency convertors and amplifiers operating over wide frequency ranges from the ultraviolet to the mid-infrared\cite{tyumenev2020narrowband,belli2015vacuum,mridha2019thresholdless,wang2020high,li20203,wang2024synthesizing}. It is worth noting that gas-based SRS and molecular modulation are nonlocal nonlinear phenomena that cannot be directly scaled using the strategy outlined in Ref.~\onlinecite{Heyl:16}, which was conceived to scale quasi-instantaneous nonlinear effects such as self-phase modulation in noble gases. Instead, we have found that it is useful to adopt a formalism similar to that employed to scale bright soliton dynamics in standard solid-core fibers~\cite{agrawal2000nonlinear}, photonic crystal fibers~\cite{RevModPhys.78.1135}, gas-filled capillaries~\cite{Travers:2019} and ARFs \cite{Schade:21}. This is based on the definition of the soliton order that depends on the ratio of two characteristic lengths related to dispersion and Kerr nonlinearity~\cite{agrawal2000nonlinear}. In our case, we will see that the ratio of specific characteristic lengths defining SRS-driven dynamics enables a direct scaling of the subsequent nonlinear evolution.

Although our scaling rules will apply to the general Eqs.~\eqref{eq:MBe_equation00} and \eqref{eq:MBe_equation0} that are valid both in transient SRS (pump pulse duration $\tau_p\leq T_2$) and steady-state regimes ($\tau_p\gg T_2$), we will consider the latter as it is more amenable to analysis. Moreover, to illustrate the procedure we will also restrict our treatment to the most relevant optical fields (Pump, Stokes and anti-Stokes), whose nonlinear interplay will be enough to construct scale-invariant dynamics under very general conditions. If the pump is not depleted (although this condition is not critical), we obtain the following equations for the coupled evolution of the first Stokes and anti-Stokes bands \cite{Boyd}:
\begin{eqnarray}
    \frac{\partial E_{1}}{\partial z} &=& -\frac{\omega_1}{2\omega_0}g_1 I_p \left(E_1 + \frac{\kappa_{1,0}}{\kappa_{1,1}}E_{-1}^*e^{-i\vartheta z}\right),\\
    \frac{\partial E_{-1}^*}{\partial z} &=& \frac{1}{2}g_0 I_p \left(E_{-1}^* + \frac{\kappa_{1,1}}{\kappa_{1,0}}E_1^*e^{i\vartheta z}\right).
\end{eqnarray}

The formal solutions of the optical fields are $E_{-1/1}(z)=E_{-1/1}(0)e^{(g_0I_p\rho \pm i\vartheta)z/2}$, where $E_{-1/1}(0)$ are the initial conditions and $\vartheta = 2\beta_0 - \beta_{-1} - \beta_{1}$ is the spatial dephasing between the interacting fields (see Fig.~\ref{FIG1}(c)). With these parameters, we can define both the dephasing length $L_D=\pi/\vartheta$ and the gain length $L_G = 1/g_0I_p$, which together give rise to the so-called gain reduction factor $\rho$~\cite{bauerschmidt2015dramatic}:
\begin{equation}
    \rho = \left|\text{Re} \sqrt{\left(\frac{q-1}{2}\right)^2-\frac{\vartheta}{g_0I_p}\left(\frac{\vartheta}{g_0I_p}+i(q+1)\right)}\right| - \frac{q-1}{2},
    \label{eq:gain_suppression}
\end{equation}
where $q=g_{1}\omega_{1}/g_0\omega_0$ remains largely invariant under pressure changes. Note that $\rho$ depends on the ratio $\vartheta/g_0I_p$, which is proportional to $L_G/L_D$\cite{Hosseinni_gainsup} in close analogy to the soliton order~\cite{agrawal2000nonlinear}. This parameter characterizes the coherent suppression of the Raman gain at $\vartheta = 0$ due to the precise balance between the creation (during P-S conversion) and annihilation  (during P-AS conversion) of optical phonons~\cite{PhysRevLett.12.504,Duncan:86}, taking values between $0$ and $1$ (see Fig.~\ref{FIG1}(d)). In other words, $\rho$ captures the dynamical competition between sidebands and can therefore be used as a ruler to reproduce the same dynamics under very different parametric landscapes. Hence, SRS dynamics will be scalable provided $\rho$ (or equivalently $L_G/L_D$) remains invariant upon changes in the physical parameters of the system. Remarkably, although Eq.~\eqref{eq:gain_suppression} has been derived under steady-state conditions, it remains largely valid in the more general transient SRS regime\cite{Frank_wise, bauerschmidt2015dramatic, chen2023femtosecond}.

Let us illustrate our approach with an example. In the steady-state SRS regime, the exponential amplification of the Stokes intensity can be approximated as $I_s \propto \exp{(g_0I_p\rho z)}$\cite{Hosseinni_gainsup}. Thus, if we scale the propagation length $z$ as $z'=\eta z$, where $\eta$ is an arbitrary scaling constant, then $L_G$ must necessarily scale as $L_G'= \eta L_G$ in order to keep their balance unaltered. Then, as $\rho$ must remain constant, this automatically implies that $L_D'=\eta L_D$. Although the scaling of the effective lengths is enough to ensure scale-invariant SRS dynamics, it is useful to obtain the scaling rules of all the relevant physical parameters in the system such as gas pressure, core radius and pump pulse energy. This is because they will be, along with the propagation length, the key ingredients defining your optical system. To do this, we first approximate the dispersion landscape of the gas-filled ARF given by Eq.~\eqref{eq:propagationctt0} using a Taylor expansion:
\begin{gather}
    \beta_l \approx \frac{\omega_l}{c}+\frac{\omega_l\delta \tilde p \tilde T }{2 c } - \frac{u_{01}^2 c}{2\omega_la^2},
    \label{eq:Taylorbeta}
\end{gather}

that permits to recast the dephasing as:
\begin{gather}
    \vartheta = \sum^1_{l=-1}\mu_l \beta_l = \tilde A \tilde p - \tilde B/a^2,
    \label{eq:sum_betas}
\end{gather}

where $\mu_l=(-1)^l(2-|l|)$, $\tilde A = \sum_l\mu_l \omega_l\delta\tilde T/2c$ and $\tilde B =\sum_l \mu_l u_{01}^2 c/2\omega_l$. Then, in order to satisfy $L_D'=\eta L_D$, the following equation must be fulfilled:
\begin{gather}
    \tilde A \left(\frac{\tilde p}{\eta}-\tilde p'\right) = \tilde B \left(\frac{1}{\eta a^2}-\frac{1}{a'^2}\right).
    \label{eq:master_eq_nobuffer}
\end{gather}

This equation is key to our analysis since it univocally defines the relation between pressure and core radius that gives rise to a scalable scenario. In particular, since both $\tilde A, \tilde B > 0$, the trivial solution of setting the terms in parentheses to zero yields $p'=p/\eta$ and $a'=\sqrt{\eta}a$. On the other hand, the condition $L_G'=\eta L_G$ can be used to scale the pump pulse energy as $\cal E'$$=f(\eta)$$\cal E$, where $f(\eta)=g_0(p)/g_0(\eta^{-1}p)$ and $g_0$ is given by Eq.~\eqref{eq:ramangain}. The scaling of the physical parameters of the system is summarized in Table ~\ref{table1}.
\begin{table}[h!] 
\centering 
\begin{tabular}{cc}
\toprule[1.5pt] \multicolumn{2}{c}{Physical parameters} \\ \midrule[1.25pt] 
Original & Scaled \\ \midrule[1.25pt]
\rowcolor[gray]{.9} $z$ & $\eta z$  \\ 
                    $a$ & $\eta^{1/2}a$  \\ 
\rowcolor[gray]{.9} $p$ & $\eta^{-1}p$\\
                    $\epsilon_p$ & $f(\eta)\epsilon_p$\\
\bottomrule[1.5pt] 
\end{tabular}
\caption{Summary of scaled variables with the $p'=p/\eta$ and $a'=\sqrt{\eta}a$ condition imposed in the steady state regime.}
\label{table1} 
\end{table}

We must emphasize that in some practical scenarios of relevance, it might be useful to find scalable dynamics when one (or more) of the physical parameters listed above is predefined (e.g., operation at a fixed pressure or core radius different from the originals). In those cases, Eq.~\eqref{eq:master_eq_nobuffer} still holds but the scaling of the different parameters with $\eta$ will not directly follow Table~\ref{table1} (for example, $\eta = 1$ will not indicate the trivial case where everything remains unchanged). We will elaborate on this exception with  examples in Sections \ref{Section A} and \ref{Section B}.

To summarize, the general workflow of our approach to obtain scalable SRS dynamics in gas-filled ARFs is the following. Since the gain reduction factor $\rho$ must be preserved, we necessarily need to follow the corresponding isopleth in Fig.~\ref{FIG1}(c) or look for an equivalent one on the opposite side of the gain suppression line $\rho = 0$. To do so, we first obtain $\vartheta'$ (Fig.~\ref{FIG1}(d)) to fix $L_D'$ and then modify the pulse energy $\cal E$ so as to obtain the required $L_G'$ that keeps $\rho' = \rho$. As discussed above, this process will enable the scalability of the dynamics governed by the set of coupled nonlinear Eqs.~\eqref{eq:MBe_equation00}–\eqref{eq:MBe_equation0} both in the steady-state and transient regimes, provided the dephasing time (or equivalently, $\Delta\nu$) is preserved.

Finally, for the comparison with the full numerical simulations using Eqs.~\eqref{eq:MBe_equation00}–\eqref{eq:MBe_equation0} we will calculate the scaled physical parameters using a custom optimization routine that uses the full dispersion (Eq.~\eqref{eq:propagationctt0}) instead of the approximate expression (Eq.~\eqref{eq:Taylorbeta}) used in the analysis outlined above.

\section{Results and discussion}

In this Section we will test the robustness of the scaling formalism within realistic experimental constraints  for the different physical parameters: $r\in[10,50]$ \textmu m, $p\in[0,100]$ bar and $\cal E$$\in[0,500]$ \textmu J. To do so, we will study the nonlinear SRS dynamics of narrowband pump pulses of different central wavelengths propagating in H$_2$-filled ARFs. In particular, we will consider two different scenarios where the zero-dispersion point (ZDP) lies either spectrally away (Section \ref{Section A}) or in the vicinity of the interacting fields (Section \ref{Section B}). Then in Section \ref{SectionC} we will discuss how gas mixtures aid the scaling of transient SRS dynamics  when the Raman-active gas pressure changes.
Finally, as a direct application of our results, in Sections \ref{Section D}, ~\ref{Section E} we will show how the molecular modulation of an arbitrary signal can be scaled. For that purpose, we will simulate the recent demonstration of highly-efficient molecular modulation of quantum light~\cite{tyumenev2022tunable}, lowering the pressure needed without losing efficiency.

\subsection{Free-space-like dynamics} \label{Section A}

\begin{figure}[h!]
\includegraphics{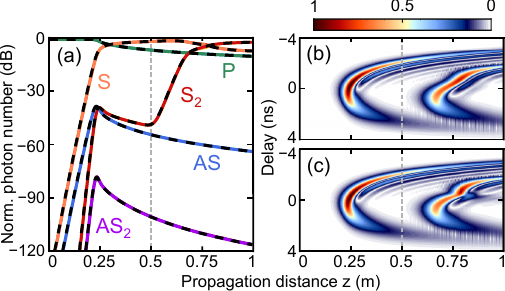}
    \caption{(a) Evolution of the normalized photon number of the 266-nm
pump (P) and its first Stokes (S), second Stokes (S$_2$), first anti-Stokes (AS) and second anti-Stokes (AS$_2$) sidebands. The simulation using data set (0.0) is represented by solid lines and that using data set (0.1) is represented by dashed lines. (b,c) Spatiotemporal evolution of the amplitude of the induced Raman coherence in the scenario characterized by (b) data set (0.0) and (c) data set (0.1). Each plot is normalized to its maximum value.}
    \label{FIG2}
\end{figure}

When the gaseous contribution to the overall ARF dispersion dominates over the geometrical one so that the influence of the core radius is negligible, the dispersion converges to that of free space and $\beta_l \approx\omega_ln_g/c$. Following Eq.~\eqref{eq:propagationctt0}, this will normally happen at very large core radii and gas pressures, being particularly acute in the ultraviolet where all interacting fields lie in the normal dispersion regime and far from the ZDP~\cite{mridha2019thresholdless}. To illustrate this regime let us consider the propagation of 3.8 ns-pump pulses of 266 nm central wavelength and $\cal E$ $= 5  $ \textmu J along an ARF filled with 30 bar H$_2$. The numerical evolution of the pump along with the first Stokes (orange-solid line), second Stokes (red-solid line), first (blue-solid line) and second (purple-solid line) anti-Stokes bands is shown in Fig.~\ref{FIG2}(a). For the scaling of these dynamics we will assume that the pressure remains constant ($p' = p$), meaning that the Raman gain is also constant and therefore both energy and core radius are trivially scaled through the intensity yielding $L_G'=L_G$. In this regime, very similar to that discussed in Ref.~\onlinecite{Heyl:16}, $z'=z$ and $L_D'=L_D$ since the overall dephasing is largely independent of the core radius $a$. The resulting parameters are summarized in Table~\ref{table2}, where (0.0) is the original data set and (0.1) is the scaled one. As we can see, when both the scaled core radius $r'$ (chosen to be larger than the original) and the pulse energy are conveniently set, $\rho$ is preserved (its near-unity value indicates that coherent gain reduction effects are absent).

\begin{table}[h!] 
\centering 
\begin{tabular}{ccccccc}
 \toprule[1.5pt] 
Set & $P_{H_2}$ (bar) & $\cal E$ (\textmu J) & r (\textmu m)& $L_D$ (mm) &$L_G$ (mm)& $\rho$\\ \midrule[1.25pt] 
\rowcolor[gray]{.9} (0.0)& 30  &5.0&20&2.6&5.8&0.98\\
         (0.1)&  30  &11.3&30&2.6&5.8&0.98\\
\bottomrule[1.5pt] 
\end{tabular} 
\hspace{0.5cm}
    \caption{Data sets of the physical parameters used in the simulations for the case of 266 nm pumping. The thickness of the capillary walls is chosen so that all interacting fields lie away from loss-inducing resonances. Data set (0.0) corresponds to the benchmarking case.}
    \label{table2}
\end{table}
  
The comparison of the evolution of the scaled data set (dashed lines) with the original one is shown in Fig.~\ref{FIG2}(a). The agreement is very good, as expected for such trivial scaling. For completeness, in Figs.~\ref{FIG2}(b)-(c) we display the spatio-temporal dynamics of the Raman coherence induced in the system for the original (b) and scaled (c) data sets. The C-shaped profile developed along the first half of the fiber corresponds to the coherence generated by the pump-first Stokes beat-note. The more complex evolution in the second half of the fiber (beyond the vertical dashed line) is related to the onset of the first-second Stokes beating. It is evident that, in this region, the dynamics is not exactly reproduced between the two data sets. The reason is the sensitivity of the second Stokes (341 nm) to the position of the ZDP (884 nm for data set (0.0) and 1079 nm for data set (0.1)), which renders our free-space-like approximation invalid. We quantified the accuracy of the scaling by calculating the root-mean-squared error (RMSE) in both regions, before and after the vertical dashed line, as well as the total error. The overall RMSE is 2\% and is approximately equal to that obtained for the second half of the evolution, meaning that the fidelity of the scaling involving discrete bands with a total spectral coverage exceeding 0.5 PHz is very high.

In this example we have seen that the proximity of the ZDP to the higher-order Stokes sidebands affects the scaling since $L_D$ only accounts for the dephasing between the pump and its first Stokes and anti-Stokes bands. In the next section, however, we will see how the SRS dynamics can also be scaled when the ZDP is very close to the interacting fields.

\subsection{Micro-confined dynamics and role of T$_2$} \label{Section B}

\begin{figure}[t]
    \centering
\includegraphics[width=\linewidth]{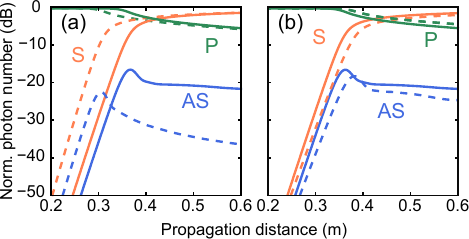}
    \caption{Evolution of the normalized photon number of the 532-nm pump (P) and its first Stokes (S) and anti-Stokes (AS) sidebands. The simulation using data set (1.0) is represented by solid lines whereas data sets (1.1) in (a) and (1.2) in (b) are represented by dashed lines.}
    \label{FIG3}
\end{figure}

When the system described before is pumped at 532 nm in the visible domain, the geometric contribution to the dispersion can no longer be neglected and will enter the scaling through $L_D$. This means that, if the radius is changed, the pressure will also have to be adjusted in order to access a scalable landscape. Hereafter we will restrict the discussion to only 3 fields (pump, first Stokes and first anti-Stokes) for the sake of simplicity. Let us now consider as benchmarking parameters those included in the data set (1.0) in Table~\ref{table3}. The evolution of 532 nm-pulses and their bands under these conditions is shown in Fig.~\ref{FIG3}(a) (solid lines). If we now increase the radius while keeping the pressure fixed, $L_G'$ can be made equal to $L_G$ through adjustment of the pulse energy, but $L_D'\neq L_D$ since the dephasing cannot be matched in this case. As a result, $\rho'\neq \rho$ (see data set (1.1) in Table~\ref{table3}) and the overall dynamics will not be scalable (see the dashed lines in Fig.~\ref{FIG3}(a)). 
Hence, as dictated by our general scaling strategy (represented by Eq.~\eqref{eq:master_eq_nobuffer}), both pressure and radius must vary. By choosing $\eta = 1$ for simplicity and keeping the core radius fixed to that used in the data set (1.1), we obtain the new data set (1.2) (see Table~\ref{table3}) using our formalism. In this case, we can see that $\rho'=\rho$, the pump energy is the same as that used in (1.1) but the pressure must be reduced to $p'$ = 13.6 bar in order to satisfy $L_D' = L_D$. The full numerical propagation using these parameters (data set (1.2)) is represented in Fig.~\ref{FIG3}(b). Although the different photon number evolution lines are much closer to the originals than those displayed in Fig.~\ref{FIG3}(a), they are still not quantitatively matching. The reason is the unmatched dephasing time $T_2$ in both data sets due to the difference in gas pressure. The nonlocal contribution of $T_2$ is particular of the transient SRS regime and was not considered in the analysis presented in either Ref.~\onlinecite{Heyl:16} or the bright soliton literature\cite{Schade:21}, where mainly quasi-instantaneous nonlinear effects are usually considered. Unfortunately, when only a pure Raman-active gas is present, there are not any additional knobs to compensate for the difference in $T_2$ (the pulse duration could mitigate its influence to some extent~\cite{hosseini2017enhanced}, but it is not easy to tune in laser systems delivering narrowband pulses). Using gas mixtures, however, can crucially enable scalable dynamics with different gas pressures as we will show in the next Section.

\begin{table}[h!] 
\centering 
\begin{tabular}{ccccccc}
 \toprule[1.5pt] 
Set & $P_{H_2}$ (bar) & $\cal E$ (\textmu J) & r (\textmu m)& $L_D$ (mm) &$L_G$ (mm)& $\rho$\\ \midrule[1.25pt] 
\rowcolor[gray]{.9} (1.0)&  20  &5&11.3&30&6&0.51\\
         (1.1)&  20  &10&16&16&6&0.69\\
         (1.2)&  13.6&10&16&30&6&0.51 \\
         \bottomrule[1.5pt] 
\end{tabular} 
\hspace{0.5cm}
    \caption{Data sets of the physical parameters used in the simulations for the case of 532 nm pumping. The thickness of the capillary walls is chosen so that all interacting fields lie away from loss-inducing resonances. Data set (1.0) corresponds to the benchmarking case.}
    \label{table3}
\end{table}

\subsection{Scaling with pressure changes. Gas mixtures} \label{SectionC}

As discussed in the previous Section, the nonlocal effects associated to the dephasing (or decoherence) time $T_2$ cannot be neglected in the transient SRS regime. Hence, in order to scale the nonlinear dynamics in this regime we must ensure that the Raman linewidth remains invariant upon scaling, something practically impossible in ARFs filled with pure Raman-active gases when the gas pressure is forced to change to compensate the dispersion. To circumvent this limitation, we propose the use of a mixture of Raman-active and noble gases (in this work, xenon) that permits simultaneous adjustment of both dispersion and $\Delta\nu$ (see Eq.~\eqref{eq:Raman_linewidth} and Fig.~\ref{FIG1}(b)). Thus, including the dispersive effects of the buffer gas pressure $p_B$ in the modal propagation constants, we obtain~\cite{hosseini2017enhanced}:
\begin{gather}
    \beta_l \approx \frac{\omega_l}{c}+\frac{\omega_l\delta_R \tilde p_R \tilde T}{2 c } +\frac{\omega_l\delta_B \tilde p_B \tilde T }{2 c} - \frac{u_{01}^2 c}{2\omega_la^2},
    \label{eq:Taylorbeta}
\end{gather}
where $\delta_{R(B)}$ is the Sellmeier relation of the Raman-active  (buffer) gas. By adding an extra condition $\Delta\nu' = \Delta\nu$ that must be satisfied to ensure scalability in pressure-varying scenarios, the formalism outlined in Section \ref{scaling principles} can be generalized to account for the influence of the buffer gas as follows:
\begin{gather}
    \vartheta = \sum^1_{l=-1}\mu_l\beta_l = \tilde A(\omega_l) \tilde p - \tilde B (\omega_l) /a^2+ \tilde C(\omega_l) \tilde p_B,
\end{gather}

where the changes with respect to Eq.~\eqref{eq:sum_betas} are $\tilde A = \sum_l\mu_l \omega_l\delta_R\tilde T/2c$ and $\tilde C = \sum_l\mu_l \omega_l\delta_B\tilde T/2c$. Then, in order to satisfy $L_D'=\eta L_D$, the following equation must be fulfilled:
\begin{gather}
\tilde A \left(\frac{\tilde p_R}{\eta}-\tilde p_R'\right) - \tilde B \left(\frac{1}{\eta a^2}-\frac{1}{a'^2}\right) + \tilde C  \left(\frac{\tilde p_B}{\eta}-\tilde p'_B\right)= 0. \label{eq:master_eq_buffer}
\end{gather}

Note that in the absence of buffering (i.e. $p_B' = 0, p_B = 0$), Eq.~\eqref{eq:master_eq_buffer} becomes Eq. ~\eqref{eq:master_eq_nobuffer}, as expected. If we do not impose any particular values for the partial pressures, the effective mode radius will be given by the following expression:
\begin{gather}
    a' = \left(\frac{1}{\eta a^2}+\frac{\tilde A}{\tilde B} \left(\tilde p'_R- \frac{p_R}{\eta}\right)+\frac{\tilde C}{\tilde B} \left(\tilde p_B' - \frac{p_B}{\eta}\right)\right)^{-1/2}.
    \label{eq:buffer_radius}
\end{gather}

\begin{figure}[h!]
    \centering
\includegraphics[width=\linewidth]{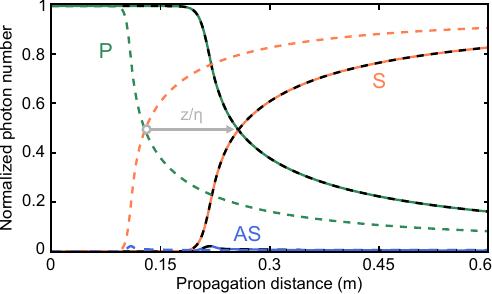}
    \caption{Evolution of the normalized photon number of the 532-nm pump (P) and its first Stokes (S) and anti-Stokes (AS) sidebands. The simulation using data set (1.0) is represented by solid lines and that corresponding to data set (1.3) is represented by color-coded dashed lines. The black-dashed line represents the results of data set (1.3) with the propagation distance rescaled by $\eta^{-1} = 2 $ to facilitate comparison. In contrast to the results shown in Fig.~\ref{FIG3}, these dynamics are displayed in linear scale for visual convenience.}
    \label{FIG4}
\end{figure}

To illustrate the performance of this generalized scaling strategy using gas mixtures, we simulated the dynamics of data set (1.0) using the same H$_2$ pressure employed in data set (1.2) (see Table.~\ref{table3}) but keeping the dephasing time unaltered. To do so, we selected a scaling parameter $\eta = 0.5$ and included $\approx 0.8$ bar of xenon to ensure $\Delta\nu' = \Delta\nu$. The simulations obtained using the parameters listed in data set (1.3) (see Table.~\ref{table4}) are displayed in Fig.~\ref{FIG4} (color-coded dashed lines). Using the conveniently prepared H$_2$ - Xe mixture, the overall benchmarking dynamics (solid lines) is precisely reproduced, albeit in a shorter propagation distance owing to the scaling of $z' = \eta z$. To facilitate comparison between the two data sets, we shifted the scaled results to twice the propagation distance (indicated with an arrow in Fig.~\ref{FIG4}) so that both original and scaled dynamics precisely overlap in the graph (black dashed lines). In sharp contrast to the dynamics obtained using data set (1.2), where the dephasing time $T_2$ was different from the original, the system is now perfectly scaled. Furthermore, our scaling strategy permits not only reproducing the same dynamics over a much shorter distance such that the system can be made more compact, but also the pump pulse energy is increased by roughly 10 times without sacrificing conversion efficiency from the pump to its sidebands, with evident practical applications in, e.g., efficient Raman amplification ~\cite{PhysRevLett.103.183902} or frequency conversion~\cite{hosseini2016generation}.

\begin{table}[h!] 
\centering 
\begin{tabular}{ccccccc}
 \toprule[1.5pt] 
Set & $P$ (bar) & $\cal E$ (\textmu J) & r (\textmu m)& $L_D$ (mm) &$L_G$ (mm)& $\rho$\\ \midrule[1.25pt] 
\rowcolor[gray]{.9} (1.0)&  20  &5&11.3&30&6&0.51\\
         (1.3)&  13.6 + 0.8 &49&20.6&15&3&0.51\\
         \bottomrule[1.5pt] 
\end{tabular} 
\hspace{0.5cm}
    \caption{Data sets of the physical parameters used in the simulations for the case of 532 nm pumping and gas mixtures. Data set (1.0) corresponds to the benchmarking case with 20 bar of pure H$_2$. Instead, the pressures indicated in the data set (1.3) correspond to H$_2$ (left value) and Xe (right value), respectively.}
    \label{table4}
\end{table}

\subsection{Gas mixtures and quantum molecular modulation} \label{Section D}

\begin{figure}[ht!]
    \centering
\includegraphics[width=\linewidth]{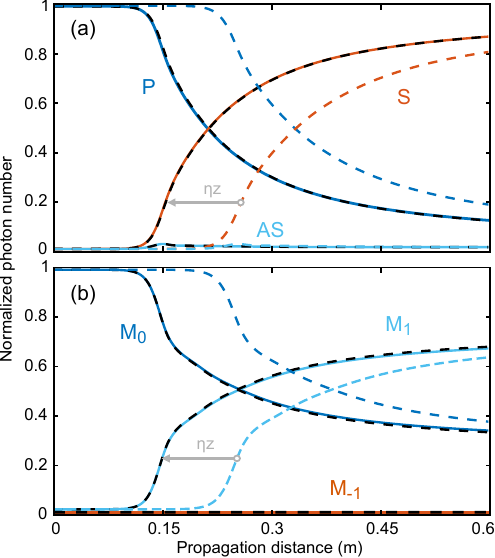}
    \caption{Evolution of the normalized photon number of (a) the 1064-nm pump (P) and its first Stokes (S) and anti-Stokes (AS) sidebands, and (b) the 1425-nm mixing (M$_0$) and its down-shifted (M$_{-1}$) and up-shifted (M$_1$) sidebands. The simulation using data set (2.0) is represented by solid lines and that corresponding to data set (2.1) is represented by color-coded dashed lines. The black-dashed lines indicate the results of data set (2.1) with the propagation distance re-scaled by $\eta^{-1} = 0.78 $ to facilitate comparison.
    }
    \label{FIG5}
\end{figure}

In this Section we will apply our scaling formalism to a more complex scenario involving the molecular modulation of an arbitrary mixing beam. In particular, we will study the recent experimental observation of highly-efficient quantum frequency up-conversion of single photons in H$_2$-filled ARF (Ref.~\onlinecite{tyumenev2022tunable}). In that work, the conversion of a mixing photon at 1425 nm to its anti-Stokes mixing band at 894 nm was reported using the Raman coherence created by 1064 nm  pump pulses, attaining efficiencies as high as 70\% at a hydrogen pressure of 70 bar. Since the technical challenges associated to operation at such high pressures can certainly impair the widespread deployment of this promising platform for lightwave quantum technologies, we will try to numerically obtain the same dynamics and conversion efficiencies reported in Ref.~\onlinecite{tyumenev2022tunable} but at much lower gas pressure. To achieve this goal, instead of using the spatial dephasing $\vartheta$ to scale the dynamics, we will consider the dephasing between the excited coherence wave $\Delta\beta$ and the corresponding mixing-upshifted transition $\vartheta_M = \Delta\beta-\Delta\beta_M$, where $\Delta\beta_M=\beta_{M_1}-\beta_{M_0}$ is the difference between the propagation constants of the mixing beam (M$_{0}$) and its up-shifted band (M$_{1}$). However, contrary to the previous scaling consideration of the dephasing $\vartheta'=\eta^{-1}\vartheta$, $\vartheta_M$ will only act as a scaling constant $\vartheta_M'=\vartheta_M$. We will impose that,  as in the experiment\cite{tyumenev2022tunable}, the mixing transition is always perfectly phase-velocity matched to the generated coherence waves ($\vartheta_M = 0$) and proceed with our scaling strategy as in the previous Section. The physical parameters used in the original experiment are summarized in the data set (2.0) of Table~\ref{table5} and the ones used in the scaled case correspond to the data set (2.1) of Table~\ref{table5}. Interestingly, it turns out that by slightly increasing the fiber core radius and pumping the system with much more energy, it is possible to obtain scalable dynamics with $\eta = 1.27$ if we use a mixture of $20$ bar of H$_2$ plus $6.2$ bar of Xe, i.e., dramatically reducing the total gas pressure from $70$ to $26.2$ bar. The comparison between the original and scaled dynamics is shown in Fig.~\ref{FIG5}, where panel (a) displays the evolution of the normalized photon number of the pump pulse and its sidebands and panel (b) displays the same information for the mixing and its sidebands. It is clear that the dynamics is very precisely reproduced using the scaled physical parameters, the small discrepancies observed being attributed to numerical rounding errors.

\begin{table}[h!] 
\centering 
\begin{tabular}{cccccccc}
 \toprule[1.5pt] 
Set & $P$ (bar) & $\cal E$ (\textmu J) & r (\textmu m)& $L_D$ (mm) &$L_G$ (mm)& $\rho$\\ \midrule[1.25pt] 
\rowcolor[gray]{.9} (2.0)& 70 + 0  & 115 & 28.5 &17&4.9&0.44\\
                    (2.1)& 20 + 6.2 &381 & 31.1&21&6.2&0.44\\
         \bottomrule[1.5pt] 
\end{tabular} 
\hspace{0.5cm}
    \caption{Data sets of the physical parameters used in the simulations describing the results reported in Ref.~\onlinecite{tyumenev2022tunable} using 1064 nm pumping. Data set (2.0) corresponds to the benchmarking case with 70 bar of pure H$_2$. Instead, the pressures indicated in the data set (2.1) correspond to H$_2$ (left value) and Xe (right value). Note that the pump energy is about 3 times higher in the scaled scenario due to a combination of Raman gain reduction and larger core radius.}
    \label{table5}
\end{table}

\begin{figure}[ht!]
    \centering
\includegraphics[width=\linewidth]{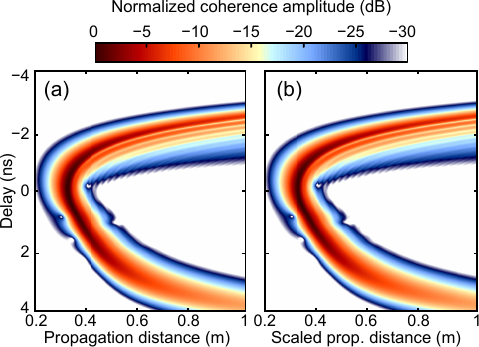}
    \caption{Simulated spatiotemporal evolution of the Raman coherence in (a) the original benchmarking case and (b) the scaled one. Note that the horizontal axis in panel (b) has been re-scaled to $\eta^{-1}=0.78$ to facilitate comparison.}
    \label{FIG6}
\end{figure}

In Fig.~\ref{FIG6} we show the spatiotemporal evolution of the induced Raman coherence, both for the original case (data set 2.0 and panel (a)) and the scaled one (data set 2.1 and panel (b)). Note that the horizontal propagation axis in (b) has been conveniently re-scaled using to $z'=1.27z$ to ease comparison. Both two-dimensional maps are normalized to their own maxima and displayed in decibel scale to enhance their differences, which are not appreciable below the $-25$ dB level. This fact, combined with a RMSE of 0.2\%, constitutes the best testimony to the fidelity of the scaling. The only quantitative difference between both scenarios is the actual coherence amplitude, much larger in the scaled simulation due to the proportionality $I_p/I_p'\propto Q/Q'$.

We must note that the experimental constraints on the physical parameters forced the choice of a scaling parameter $\eta > 1$, implying that the scaled dynamics develops on a longer propagation distance. In the next Section we will discuss how the precise tuning of the capillary-wall thickness can help to adjust $\eta$ without compromising the performance of the system.

\subsection{Dispersive fine-tuning using capillary-wall resonances} \label{Section E}

So far we have only considered the resonance-free dynamics of the interacting fields, i.e. when they are spectrally far away from any loss-inducing anti-crossings between core-guided light and modes localized in the glass walls of the tubular cladding~\cite{wei2017negative}. However, if any of the propagating fields lies close to these resonances, their dispersive influence on the propagation constant is not negligible and must be considered. In this Section we will show how using this effect can help to further adjust $L_D$, enabling, for instance, access to $\eta$ values smaller than that used in the previous Section while keeping the scaled dynamics largely unaffected.

When the resonances are relevant, the hollow capillary model (Eq.~\eqref{eq:propagationctt0}) can be generalized to account for their influence as follows\cite{zeisberger2017analytic}:
\begin{equation}
    \beta= k_0n_g - \frac{u^2_{01}}{2k_0n_ga^2}-\frac{u_{01}^2 \cot\phi (n'+1)}{2k_0^2n_g^2a^3(n'+1)^{1/2}},
    \label{eq:propagationctt}
\end{equation}
where $\phi=k_0d(n_w^2-n_g^2)^{0.5}$, $d$ is the average capillary-wall thickness, $n_w(\omega)$ is the refractive index of the silica capillary walls and $n'=(n_w/n_g)^2$.

\begin{figure}[ht!]
    \centering
\includegraphics[width=\linewidth]{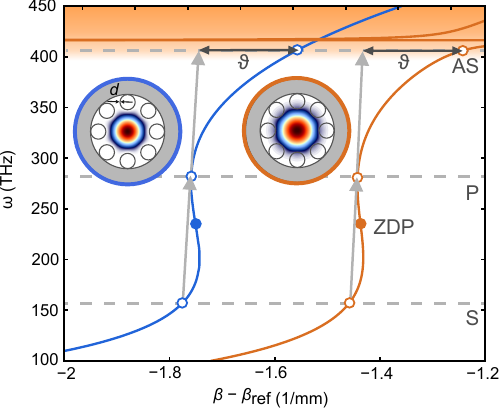}
    \caption{Dispersion diagram of a gas-filled ARF in the vicinity of the ZDP. The blue line was obtained from the data set (2.0) and the orange from the data set (2.1) with a modified $d = 340$ nm. The linear function $\beta_{ref}$ is subtracted from the propagation constants to bring out the underlying S-shaped profile. Pump (P), Stokes (S) and anti-Stokes (AS) frequencies are indicated with horizontal dashed lines and white disks at the crossing points with the curves. ZDPs are indicated by full disks. The grey arrows represent the coherence waves so that their separation from the actual dispersion curves indicate the dephasing $\vartheta$. The high-loss region is orange-shaded. Insets: Fundamental core mode calculated using finite-element modelling and overlaid onto the fiber structure for (left) the original data set (2.0) with $d = 270$ nm and (right) the data set (2.1) with $d = 340$ nm.
    }
    \label{FIG7}
\end{figure}

The capillary-wall thickness $d$ (see the insets in Fig.~\ref{FIG6}) determines the spectral position of the resonance anti-crossings. Thus, when the cladding modes are phase velocity-matched with the core modes, their propagation constant is modified and confinement is lost. The influence of these resonances on various nonlinear optical processes has already been extensively studied \cite{sollapur2017resonance,Tani:18,Chen:20,deng2023midinfraredpulse}. Here we propose to use them as an additional knob to enable scalable SRS dynamics under different conditions. As long as their influence is not very strong (i.e., when they are not spectrally close enough to cause high loss and uncontrollable dispersion changes), we can utilize the asymptotic slope variation produced by the last term in Eq.~\eqref{eq:propagationctt} to modify the propagation constants of specific spectral bands while leaving the others untouched. In the quantum molecular modulation case discussed in Section \ref{Section D}, $d = 270$~nm, $L_D = 1.27L_D'$ and this proportionality could not be made closer to unity due to the impossibility to independently adjust $\vartheta_M$ and $\vartheta$. We can, however, effectively decouple them by bringing the first-order resonance closer to the anti-Stokes band of the 1064 nm pump by using an ARF with $d \approx 340$ nm and the same core diameter. As shown in Fig.~\ref{FIG7}, this would enable, for instance, matching the dephasing $\vartheta$ in both data sets (2.0) and (2.1) for the simulations depicted in Fig.~\ref{FIG5}. This is because the near-resonance mode expands more into the glass region (see the calculated mode profiles displayed in the insets of Fig.~\ref{FIG7}), rising its effective index. Moreover, under these circumstances we could access $\eta = 1$ simply by increasing the pump pulse energy used in (2.1) to 480 \textmu J. Of course, bringing the resonance closer to the anti-Stokes band also increases its attenuation to $\sim 8$ dB/m, a tolerable value since SRS-related experiments in gas-filled ARFs typically require fiber lengths below 1 m \cite{mridha2019thresholdless,tyumenev2020narrowband}.

\section{Conclusions and outlook}

We have introduced a robust scaling strategy for nonlinear SRS dynamics in gas-filled ARFs. From the basic governing equations, we have obtained general scaling rules based on the preservation of the gain reduction factor $\rho$ or, equivalently, the ratio of its associated gain ($L_G$) and spatial dephasing ($L_D$) lengths. We tested our formalism under very different conditions, namely when the system is pumped very far from the ZDP (Section \ref{Section A}) or in its vicinity (\ref{Section B}), obtaining very good agreement even when the dynamics involved optical fields with a maximum spectral separation of 0.5 PHz. The discrepancies observed when the Raman-active gas pressure changes as a result of the scaling procedure were attributted to the variation of the dephasing time $T_2$, and could be mitigated using gas mixtures to preserve the linewidth. Finally, we tested our results in the real scenario of a quantum molecular modulator for single photons~\cite{tyumenev2022tunable}, showing that, by judicious addition of a buffer gas and adjustment of the physical parameters (i.e. pulse energy, core radius and capillary-wall thickness) within realistic bounds, it would be possible to dramatically reduce the required gas pressure from 70 to 26.2 bar without losing performance. This example clearly highlights the potential of our results to boost the conception of novel fiber-based devices based on SRS for the efficient generation, amplification or frequency conversion of classical and quantum light under very general conditions.

\begin{acknowledgments}
This work was supported by the grants PID2021-123131NA-I00, PID2021-122505OBC31, PRE2022-102843 and TED2021-129959B-C21, funded by MICIU/AEI/10.13039/501100011033, by “ERDF a way of making Europe”, by the “European Union NextGenerationEU/PRTR” and "ESF+", and the Gobierno Vasco/Eusko Jaurlaritza (IT1452-22), ELKARTEK ($\mu$4Smart-KK-2023/00016 and Ekohegaz II-KK-2023/00051), and the “Translight” initiative of UPV/EHU, and the IKUR Strategy of the Department of Education of the Basque Government through the grant IKUR\_IKA\_23/03. M. S.-H. acknowledges support from the predoctoral grant "Formación de Profesorado Universitario" FPU22/01451 from the Spanish Ministry of Science, Innovation and Universities (MICIU). PA acknowledges support from the Basque Government grant for predoctoral researchers, Ref. PREGV23/47.
\end{acknowledgments}

\section*{AUTHOR DECLARATIONS}
\subsection*{Conflict of interest}
The authors have no conflicts to diclose.

\section*{DATA AVAILABILITY STATEMENT}
The data that support the article are available within the article.

\section*{References}
\bibliography{sample}

\end{document}